\documentclass[a4paper]{mem}
\usepackage{natbib}
\usepackage{graphicx}
\usepackage[a4paper]{hyperref}
\idline{73}{1}
\begin{document}
   \title{Globular cluster abundances in the light of 
3D hydrodynamical model atmospheres
}

   \author{Martin Asplund
}


   \institute{Research School of Astronomy and Astrophysics,
              Mt Stromlo Observatory, Cotter Road, Weston ACT 2611, Australia 
\email{martin@mso.anu.edu.au}
             }

   \abstract{The new generation of 3D hydrodynamical model atmospheres have been
employed to study the impact of a realistic treatment of stellar convection on
element abundance determinations of globular cluster stars for a range of atomic and 
molecular lines. Due to the vastly different temperature structures in the optically
thin atmospheric layers in 3D metal-poor models compared with 
corresponding hydrostatic 1D models, some
species can be suspected to be hampered by large systematic errors in existing
analyses. In particular, 1D analyses based on minority species and 
low excitation lines may overestimate the abundances
by $>0.3$\,dex. Even more misleading may be the use of molecular lines 
for metal-poor globular clusters. However, the prominent 
observed abundance (anti-)correlations
and cluster variations are largely immune to the choice of model atmospheres. 
   }
   \authorrunning{Martin Asplund}
   \titlerunning{Globular cluster abundances and 3D model atmospheres}
   \maketitle
%

\section{Introduction}

Determining stellar element abundances play a crucial role in 
the efforts to improve our understanding of formation and evolution 
of globular clusters.
The term {\em observed abundances} is somewhat of
a misnomer however, since the chemical composition can not 
be inferred directly from an observed spectrum.
The obtained stellar abundances are therefore never more 
trustworthy than the models of the stellar atmospheres and
the line formation processes employed to analyse the observations.
Traditionally, abundance analyses of late-type stars rely on a number of 
assumptions, several of which are known to
be of quite questionable nature. In standard analyses, 
the employed model atmospheres are
one-dimensional (1D, either plane-parallel or spherical), time-independent,
static and assumed to fulfull hydrostatic equilibrium. 
Energy transport by convection is approximated by the rudimentary
mixing-length theory while otherwise radiative equilibrium is enforced. 
Furthermore, local thermodynamic equilibrium (LTE) is normally assumed
both for the construction of the model atmospheres and in the
spectrum synthesis.   
It should come as no surprise that abundance analyses performed along
these lines may well contain significant systematic errors due
to the adopted simplifications and approximations. 

Perhaps the most severe shortcoming in standard analyses is the
treatment of convection. 
For late-type stars, the surface convection zone reaches
the stellar atmosphere, which thereby directly affects the 
emergent spectrum. 
The solar granulation is the observational
manifestation of convection: concentrated, rapid,
cold downdrafts in the midst of broad, slow, warm upflows. 
Qualitatively similar granulation properties are expected
in other solar-type stars, as indeed confirmed by
3D numerical simulations (e.g. Nordlund \& Dravins 1990;
Asplund et al. 1999; Asplund \& Garc\'{\i}a P{\'e}rez 2001;
Allende Prieto et al. 2002)
and indicated by observed spectral line asymmetries.
The up- and downflows have radically different
temperature structures (Stein \& Nordlund 1998),
which can not be approximated by normal theoretical
1D hydrostatic model atmospheres with different
effective temperatures $T_{\rm eff}$ (Fig. 1).
Because of the photospheric inhomogeneities
and the highly non-linear and non-local nature
of spectrum formation, it is clear that no single 
1D model can be expected to properly describe all aspects of
what is inherently a 3D phenomenon
(e.g. Asplund et al. 2003b).

Here I will describe recent
progress in developing 3D hydrodynamical
model atmospheres of late-type stars and their applications to 
stellar abundance analyses, in particular for 
elements relevant for globular cluster studies.

\section{3D hydrodynamical model atmospheres}

The 3D model atmospheres which form the basis of
the abundance analyses presented here have been
computed with a 3D, time-dependent, compressible, explicit,
radiative-hydrodynamics code developed to
study solar and stellar surface convection
(Stein \& Nordlund 1998). The hydrodynamical
equations for conservation of mass, momentum
and energy are solved on a Eulerian
mesh with gridsizes of $\approx 100^3$ with 
explicit time-integration. The physical
dimensions of the grids are sufficiently large
to cover many ($>10$) granules simultaneously 
in the horizontal direction and about 13 pressure
scale-heights in the vertical. In terms of
continuum optical depth the simulations extend
at least up to log\,$\tau_{\rm Ross} \approx -5$ which for
most purposes are sufficient to avoid 
numerical artifacts of the open upper boundary on
spectral line formation. 
The lower boundary is located at large depths to
ensure that the inflowing gas is isentropic and featureless, 
while periodic horizontal boundary conditions are employed.

\begin{figure*}
\centering
\resizebox{\hsize}{!}{\includegraphics{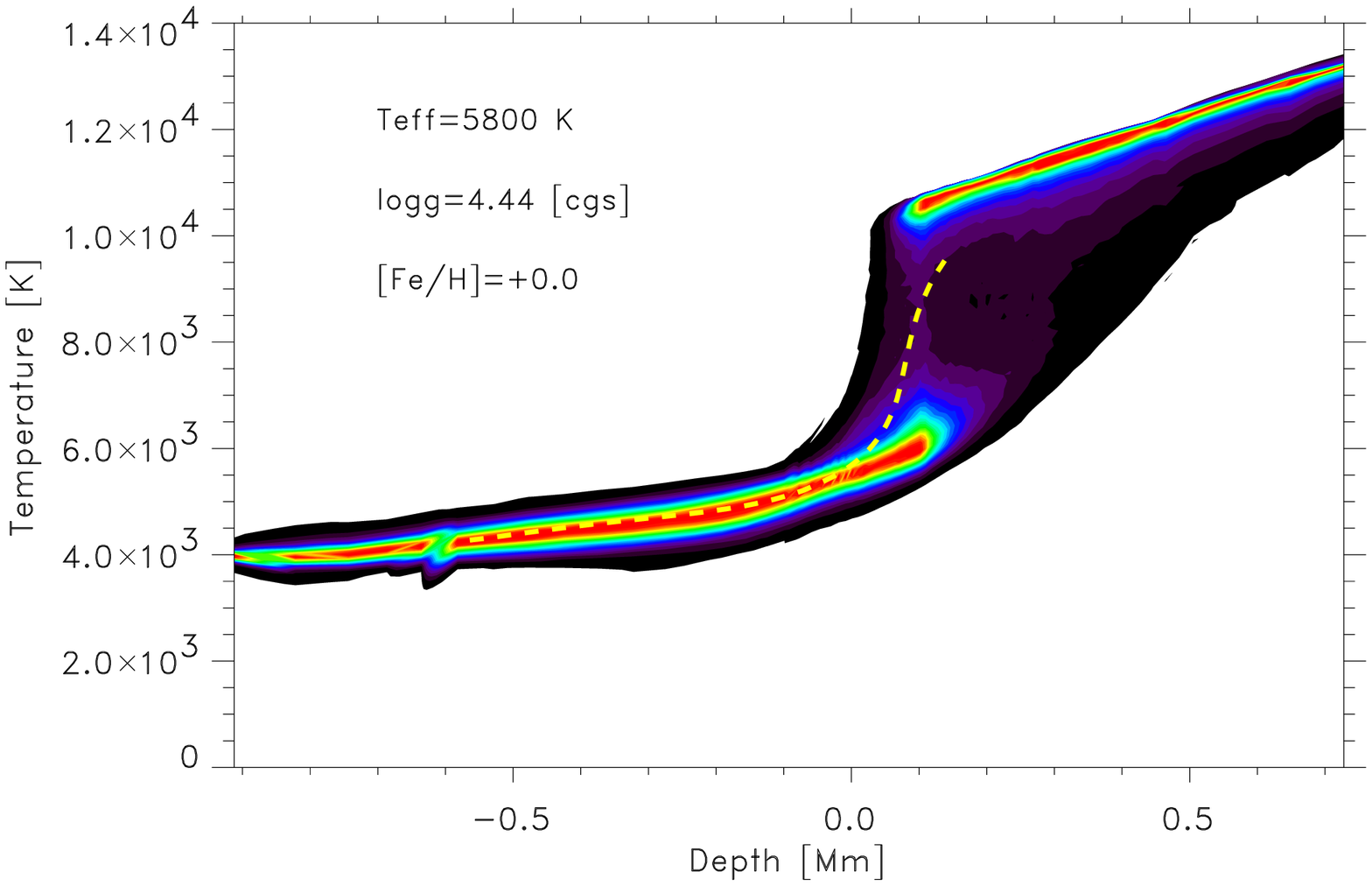}}
\resizebox{\hsize}{!}{\includegraphics{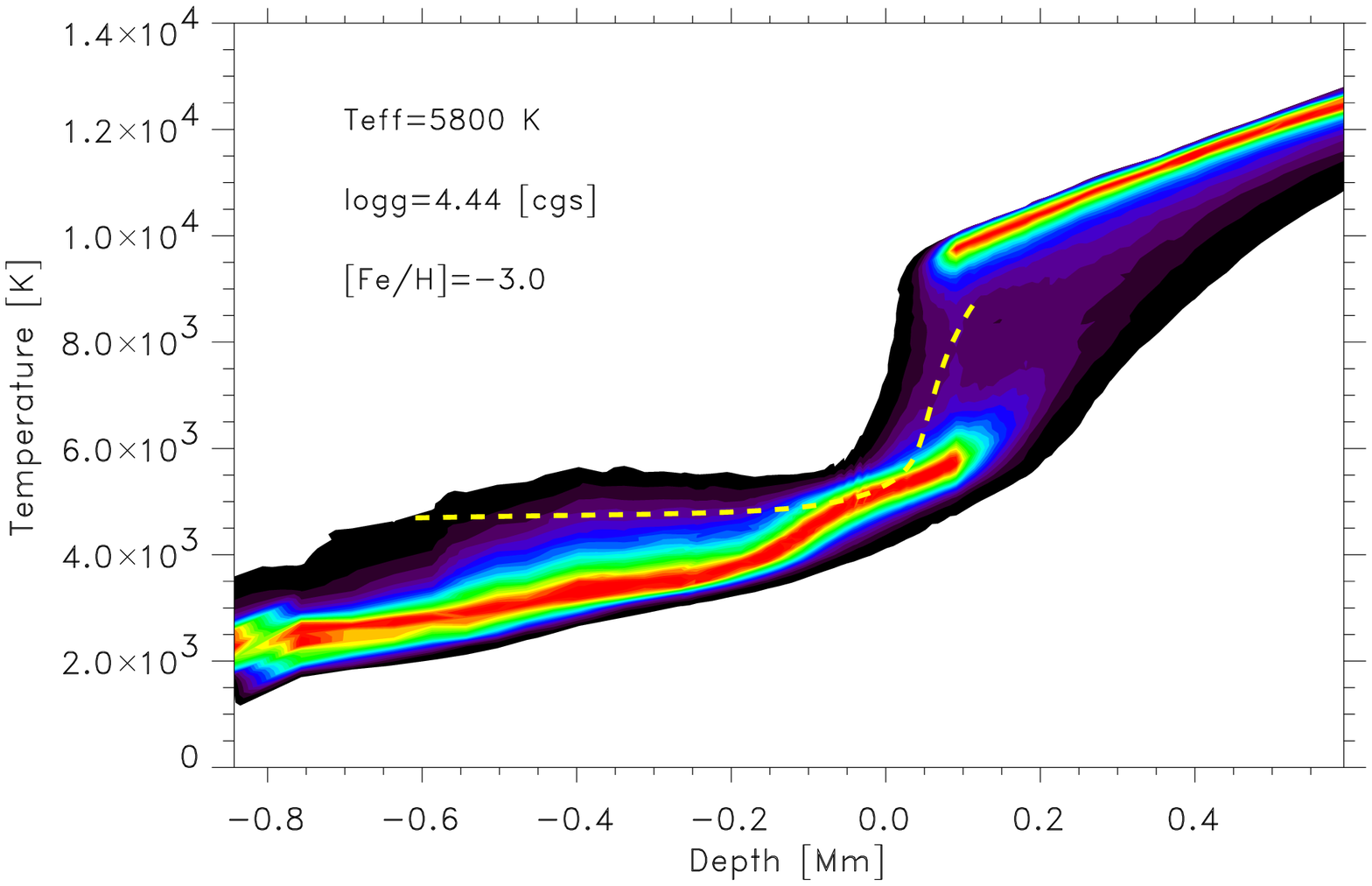}}
\caption{The resulting temperature distribution in the upper part
of 3D hydrodynamical convection simulations of the Sun ({\em upper panel})
and a metal-poor ([Fe/H]\,$=-3$) Sun ({\em lower panel}).
Also shown are the predictions from the corresponding 
theoretical 1D {\sc marcs} model atmospheres (dashed lines).
The zero-point for the depth-scale corresponds roughly to the
continuum optical depth unity. 
Note the much cooler 3D temperatures in the optically thin layers
in the metal-poor star but the rough agreement at solar metallicity
(see text for discussion).
}
\label{f:tz}
\end{figure*}

In order to obtain a realistic atmospheric structure,
it is crucial to have the best possible input physics, and
properly account for the energy exchange between
the radiation field and the gas. 
The adopted equation-of-state is that of Mihalas et al. (1988),
which includes the effects of ionization, excitation
and dissociation. The continuous opacities come from
the Uppsala package (Gustafsson et al. 1975 and subsequent
updates) while the line opacities are from Kurucz (1998, private communication).
The 3D radiative transfer is solved at each time-step
under the assumptions of local
thermodynamic equilibrium (LTE, $S_\lambda = B_\lambda$)
and opacity binning (Nordlund 1982). 
The assignment of the original 2748 wavelength points into
the different opacity bins follows from detailed monochromatic radiative
transfer calculations of the 1D averaged atmospheric structure.
The opacity binning thus includes the effects of line-blanketing
in a manner reminiscent of opacity distribution functions.

It is important to realise that {\em the simulations contain
no free parameters which are tuned to improve the agreement
with observations}. The adoption of the numerical and physical
dimensions of the simulation box is determined by practical
computional time considerations, the need to resolve
the most important spatial scales and the wish to place the artificial
boundaries as far as possible from the region of interest.
It has been verified that the resulting atmospheric
structures are insensitive to the adopted effective viscosity
at the current highest affordable numerical resolution
(Asplund et al. 2000a). The input parameters 
discriminating different models are the surface gravity log\,$g$,
metallicity [Fe/H] and the entropy of the inflowing material
at the bottom boundary. The effective temperature of
the simulation is therefore a property which depends on the
entropy structure and evolves
with time around its mean value
following changes in the granulation pattern.

The 3D hydrodynamical model atmospheres described above have sofar
been performed for solar-type main sequence and subgiant stars 
(e.g. Nordlund \& Dravins 1990; Asplund et al. 1999; 
Asplund \& Garc\'{\i}a P{\'e}rez 2001; Allende Prieto et al. 2002).
The most profound differences with predictions from theoretical 1D
hydrostatic models occur for low-metallicity stars, as first
shown by Asplund et al. (1999). In 1D, the presence of spectral lines
causes surface cooling and backwarming (e.g. Mihalas 1978), which 
translates to a shallow temperature gradient for metal-poor stars. 
In reality however, the radiative equilibrium which is 
enforced in 1D model atmospheres is not necessary fulfilled. 
Instead the temperature 
in the optically thin atmospheric layers is determined by a competition
between two opposing effects: {\em radiative heating} by absorption
of photons in spectral lines and {\em expansion cooling} of overshooting
upflowing material due to the density stratification. At solar
metallicity the two effects nearly balance, leaving the temperature
close to the radiative equilibrium expectations. At low metallicities,
however, the lack of lines produces much less radiative heating and
consequently the resulting temperature is much below the radiative 
equilibrium value (Asplund et al. 1999), as shown in Fig. \ref{f:tz}. 
At [Fe/H]\,$=-3$, the difference between 1D and 3D predictions can
exceed 1000\,K, which obviously can have a dramatic impact on
spectral features sensitive to those cool layers, such as
molecular and low excitation lines as well as minority species.

Further details of the 3D hydrodynamical model atmospheres
are available in Stein \& Nordlund (1998), Asplund et al. (1999, 2000a,b)
and Asplund \& Garc\'{\i}a P{\'e}rez (2001).

\section{3D spectral line formation}

The 3D hydrodynamical model atmospheres described in the previous
section form the basis for the 3D spectral line formation
calculations. 
The original 3D hydrodynamical simulations which includes
deep, extremely optically thick layers are interpolated to
a finer depth scale which extends only down to 
log\,$\tau_{\rm Ross} \approx 2.5$
prior to the spectral line calculations for improved numerical accuracy. 
At the same time, the simulations are interpolated to a coarser
horizontal grid to ease
the computational burden. For abundance analysis purposes this
procedure introduces non-noticable differences ($\le 0.01$\,dex
for individual snapshots and less for temporal averages).
The concepts of micro- and macroturbulence, which are
introduced in 1D analyses in order to account for the missing
line broadening, are not necessary in 3D calculations with
a self-consistent accounting of Doppler shifts arising from
the convective motions.

\begin{figure*}[t]
\centering
\resizebox{\hsize}{!}{\includegraphics{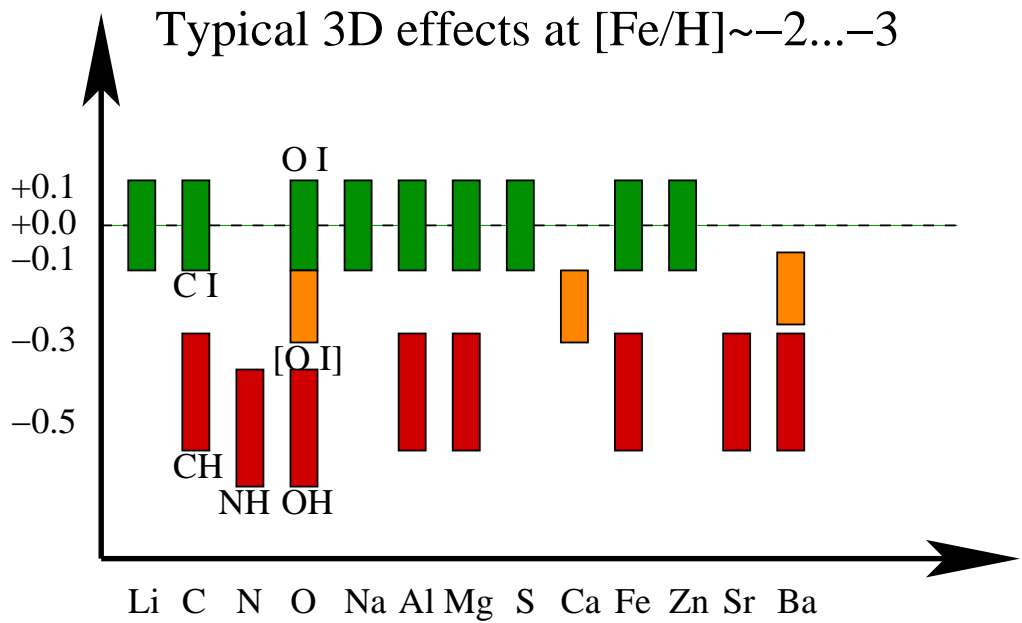}}
\caption{Schematic representation of the expected 3D LTE effects 
for different elements for $T_{\rm eff}=5800-6200$\,K stars with
[Fe/H]\,$\simeq -2$; although not quantitatively investigated here
stars with different parameters are expected to show qualitatively
similar 3D effects. While many elements apparently are insensitive
to the employment of 3D model atomspheres, some species can be
quite vulnerable due to their temperature sensitivity, 
in particular molecular lines (e.g. CH, NH, OH)
and low excitation lines. For example, while standard 1D analyses yield
accurate results for Fe\,{\sc ii} lines, Fe\,{\sc i} lines tend
to significantly overestimate the Fe abundances. Note that these
estimates are based on 3D LTE calculations, while several species
may well show significant departures from LTE. 
}
\label{f:summary}
\end{figure*}

For the spectral line formation calculations presented here the
simplifying assumption of LTE has been made, implying that the
level populations are determined by the Saha and Boltzmann 
distributions. With the source function thus known it is then
straightforward, albeit computationally intensive, to solve 
the 3D radiative transfer equation for a number of simulation
snapshots before spatial and temporal averaging of the resulting
flux profiles. 
All in all, a single temporally and spatially averaged flux profile in 3D
correspond most of the time to
$N_{\rm t}*N_{\rm x}*N_{\rm y}*N_{\rm angles}*N_\lambda \ga 10^8$
1D radiative transfer calculations.
In addition, each 3D profile is normally computed for 
at least three different abundances to enable interpolation
to the requested line strength.
Even then, such 3D LTE line calculations are achievable
on current workstations thanks to efficient numerical algorithms.
Recently, methods enabling even detailed 3D non-LTE
line formation for large model atoms have been designed
(Botnen \& Carlsson 1999) and applied to 
abundance analyses for Li 
(Kiselman 1997: Uitenbroek 1998; Asplund et al. 2003a) 
and O (Asplund et al. 2003b). 
The near future will doubtless see many more such studies
for more elements and 3D model atmospheres.

In order to investigate the possible impact of the new generation
of 3D hydrodynamical model atmospheres on globular cluster research,
I have selected a number of species and lines which are often
used to infer the chemical compositions of globular cluster stars:
Li\,{\sc i}, C\,{\sc i}, O\,{\sc i}, [O\,{\sc i}], Na\,{\sc i}, Mg\,{\sc i}, 
Al\,{\sc i}, S\,{\sc i}, Ca\,{\sc i}, Fe\,{\sc i}, Fe\,{\sc ii}, 
Zn\,{\sc i}, Sr\,{\sc ii}, Ba\,{\sc ii},
CH, NH, and OH lines. 
The 3D LTE line formation calculations were performed for solar-like
($T_{\rm eff} \simeq 5800$\,K, log\,$g = 4.4$) and turn-off stars
($T_{\rm eff} \simeq 6200$\,K, log\,$g = 4.0$) of varying
metallicities ([Fe/H]\,$ = 0.0, -1.0, -2.0, -3.0$), as well as for
a few 3D models corresponding to specific stars (Procyon, HD\,140283,
HD\,84937, G64-12). 
Unfortunately, no 3D models are yet available for giants although
work towards achieving this goal is ongoing. 
As mentioned above, no microturbulence enters
these 3D calculations. To facilitate an estimation of the 3D effects
on derived abundances, corresponding calculations were carried out
with the same code using 1D {\sc marcs} model atmospheres (Asplund et al. 1997),
adopting in all cases $\xi_{\rm turb} = 1.0$\,km\,s$^{-1}$.

\section{Results and discussion}

Fortunately, several elements appear little, if at all, affected by
the employment of 3D hydrodynamical model atmospheres instead of
classical 1D hydrostatic models, as shown in Fig. \ref{f:summary}.
In particular, lines of Li\,{\sc i} (in 3D non-LTE), 
C\,{\sc i}, O\,{\sc i}, Na\,{\sc i}, 
S\,{\sc i}, Fe\,{\sc ii}, and Zn\,{\sc i}, as well as several
Mg\,{\sc i} and Al\,{\sc i} lines, are essentially immune to the choice
of model atmospheres ($\Delta {\rm log} \epsilon < 0.1$\,dex).

The bad news, however, is that as expected in metal-poor star minority species
(e.g. Fe\,{\sc i}), low excitation 
(e.g. Mg\,{\sc i} 517\,nm, Al\,{\sc i} 394\,nm, 
Sr\,{\sc ii} 421\,nm, Ba\,{\sc ii} 455\,nm) and
all molecular lines tend to significantly over-estimate the abundances
by $\ge 0.3$\,dex when relying on 1D model atmospheres.
The reason for these differences is the much lower temperatures 
in the optically thin layers at low metallicities.
It should be noted, however, that the use of 3D models have
recently caused a significant downward revision even of the solar
C, N and O abundances (e.g. Asplund et al. 2003b) due mainly
to the presence of temperature inhomogeneities in 3D. 

As a consequence of the above-mentioned findings, it is clear that
globular cluster metallicities should be based on Fe\,{\sc ii} lines
(see also Kraft \& Ivans 2003). 
In addition, the observed O-Na, Al-Mg, C-N anti-correlations are robust
features of globular clusters as are the large cluster abundance variations. 
While absolute abundances based on molecular lines can be highly
misleading, isotopic abundance ratios are not significantly affected.

It should be emphasized, however, that the above estimates are based only
on 3D LTE calculations. In some cases there may
also be significant 3D non-LTE effects, which may diminish the overall
net effects of the new 3D model atmospheres for some species (e.g. Li\,{\sc i}, 
Asplund et al. 2003a) while aggrevate them for others 
(e.g. O\,{\sc i}, Asplund et al. 2003b). Detailed 3D non-LTE calculations
for additional elements can be expected for the future, which would
finally place the abundance analyses of late-type stars on a firm footing.

\begin{acknowledgements}
The author gratefully acknowledges financial support from
Swedish and Australian Research Councils. 
\end{acknowledgements}

\bibliographystyle{aa}

\end{document}